\documentclass[twocolumn]{article}

\usepackage{graphicx}
\usepackage{times}

\begin{document}

\title{Intergalactic dust and its photoelectric heating}
\author{Akio K. Inoue\thanks{College of General Education, Osaka Sangyo
University, 3-1-1, Nakagaito, Daito, 574-8530 Osaka: 
akinoue@las.osaka-sandai.ac.jp}, \& Hideyuki Kamaya\thanks{Department of
Earth and Ocean Sciences, National Defense Academy of Japan, Hashirimizu
1-10-20, Yokosuka 239-8686, Kanagawa}}
\date{}
\maketitle
\begin{abstract}
We have examined the dust photoelectric heating in the intergalactic
medium (IGM). The heating rate in a typical radiation field of the IGM 
is represented by $\Gamma_{\rm pe} = 1.2\times10^{-34}$ erg s$^{-1}$ 
cm$^{-3}$ $({\cal D}/10^{-4})$ $(n_{\rm H}/10^{-5}~{\rm cm^{-3}})^{4/3}$
$(J_{\rm L}/10^{-21}~{\rm erg~s^{-1}~cm^{-2}~Hz^{-1}~sr^{-1}})^{2/3}$
$(T/10^4~{\rm K})^{-1/6}$, where ${\cal D}$ is the dust-to-gas mass
ratio, $n_{\rm H}$ is the hydrogen number density, $J_{\rm L}$ is the
mean intensity at the hydrogen Lyman limit of the background radiation,
and $T$ is the gas temperature, if we assume the new X-ray photoelectric
yield model by Weingartner et al.~(2006) and the dust size distribution
in the Milky Way by Mathis, Rumpl, \& Nordsieck (1977). This heating rate
dominates the HI and HeII photoionization heating rates when the
hydrogen number density is less than $\sim10^{-6}$ cm$^{-3}$ if 
${\cal D}=10^{-4}$ which is 1\% of that in the Milky Way, although the
heating rate is a factor of 2--4 smaller than that with the old yield
model by Weingartner \& Draine (2001). The grain size distribution is
very important. If only large ($\ge0.1$ $\mu$m) grains exist in the IGM,
the heating rate is reduced by a factor of $\simeq5$. Since the
dust heating is more efficient in a lower density medium relative to the
photoionization heating, it may cause an inverted temperature--density
relation in the low density IGM suggested by Bolton et
al.~(2008). Finally, we have found that the dust heating is not very
important in the mean IGM before the cosmic reionization.
\end{abstract}


\section{Introduction}

Dust grains are formed at the end of the stellar life; 
in the stellar wind of asymptotic giant branch stars
(e.g., Ferrarotti \& Gail 2006), in the stellar ejecta of supernovae
(e.g., Nozawa et al.~2003, Rho et al.~2008), and so on.
Some of the grains grow in molecular clouds (e.g., Draine 1990). 
Some of them are destroyed by the interstellar shock 
(e.g., Williams, et al. 2006). Some of them may go out from the parent
galaxy and reach the intergalactic medium (IGM) (e.g., Aguirre et
al.~2001a,b).

The IGM is the medium between galaxies, and it occupies almost the whole
volume of the Universe. The mean density of the IGM is as low as
$10^{-7}$--$10^{-4}$ cm$^{-3}$. As found by Gunn \& Peterson (1965), 
the IGM is highly ionized after the cosmic reionization epoch 
(the redshift $z\simeq6$--10; Loeb \& Barkana 2001, Fan, Carilli, \&
Keating 2006). Thus, its temperature is $\sim10^4$ K. The IGM is filled
with the ionizing ultra-violet (UV) and X-ray background radiation which
is produced by QSOs and galaxies (e.g., Haardt \& Madau 1996).

A significant amount of metals is found in the IGM (e.g., Aguirre et
al.~2001c). Multiple supernova explosions (SN) caused by an active
star-formation in galaxies can eject the metal elements to the IGM. 
However, Ferrara, Pettini, \& Shchekinov (2000) showed that the metal
enrichment of the IGM by SN explosions is limited in relatively small
regions around star-forming galaxies and an additional physical
mechanism is required to explain the observed global enrichment of
metals in the IGM. Dust grains expelled from galaxies by the radiation
pressure due to stellar light and by the galactic wind due to multiple
SNe may contribute to the metal enrichment in the IGM (e.g., Aguirre et
al. 2001a,b). Bianchi \& Ferrara (2005) showed that relatively large
($>0.1$ $\mu$m) dust grains are not completely destroyed and reach a
significant distance (a few $\times100$ kpc) although the amount of this
intergalactic dust is too small to make a detectable extinction. 

Infrared (IR) emission from dust grains in the IGM surrounding edge-on
galaxies has been already detected (e.g., Alton, Davies, \& Bianchi
1999, Bendo et al.~2006). Moreover, IR emission from dust in the
IGM accumulated from the distant Universe may affect the cosmic
far-infrared background and the cosmic microwave background (Aguirre \&
Haiman 2000). Emission signature from dust even at the epoch of the
cosmic reionization may be detectable with a future satellite
observing the cosmic microwave background (Elfgren, D{\'e}sert, \&
Guiderdoni 2007).

Xilouris et al.~(2006) found a significant reddening of galaxies
behind the M81 group IGM which is detected by HI 21 cm emission (e.g.,
Yun, Ho, \& Lo 1994). Their measurements imply the dust-to-gas ratio in
the M81 group IGM is a factor of 5 larger than that in the Milky
Way. Such a large amount of dust in the IGM may be ejected from M82 by
its intense starburst activity (Alton et al.~1999).

Dust in the IGM affects results from the precision cosmology. 
Indeed, high redshift SNe Ia are dimmed by dust in the IGM, 
and then, the observational estimate of the distance to them 
and cosmological parameters become ambiguous 
(Goobar, Bergstr{\"o}m, \& M{\"o}rtsell 2002). 
Furthermore, future investigations of the 'equation of state' of the
Dark energy will be affected by the extinction of the intergalactic dust
even if its amount is too small to affect the conclusion of the presence
of the Dark energy (Carasaniti 2006, Zhang \& Corasaniti 2007).
 
Dust in the IGM also affects the thermal history of the IGM. In the
intracluster medium, dust grains work as a coolant because they emit
energy obtained from gas particles collisionally as the thermal IR
radiation (Montier \& Giard 2004). Such emission from some nearby galaxy
clusters can be detectable with the current and future satellites for the
IR observations (Yamada \& Kitayama 2005). Dust grains in the IGM also
work as a heating source via the photoelectric effect (Nath, Sethi \&
Shchekinov 1999). Inoue \& Kamaya (2003, 2004) proposed a possibility to
obtain an upper limit of the amount of the intergalactic dust based on
the thermal history of the IGM with the dust photoelectric heating. 

In this paper, we revisit the effect of the dust photoelectric heating
in the IGM. Recently, Weingartner et al.~(2006) revised the model of the
photoelectric yield of dust grains. They included a few new physical
processes; the photon and electron transfer in a grain, the
photoelectron emission from the inner shells of the constituent atoms of
grains, the secondary electron emission, and the Auger electron
emission. These new features reduce the photoelectric yield for moderate
energy photons of $\sim100$ eV but enhance the yield for high energy
photons of $>1$ keV. In particular, we explore the effect of the new
yield model on the photoelectric heating by the intergalactic dust in
this paper.
 
The rest of this paper consists of four sections; in \S2, we describe
the model of the photoelectric effect. In \S3, we compare heating rates
of the photoelectric effect with those of the photoionization in the
IGM. In \S4, we discuss the implications of the results of \S3. Final
section is devoted for our conclusions.

\section{Dust photoelectric effect}

\subsection{Grain charging processes}

To examine the photoelectric effect, we must specify the charge of
grains which is given by the following equation 
(Spitzer 1941, Draine \& Salpeter 1979):
\begin{equation}
 \frac{dZ_{\rm d}}{dt}=\sum_i R_i + R_{\rm pe} \,,
\end{equation}
where $Z_{\rm d}$ is the grain charge in the electron charge unit, $R_i$
is the collisional charging rate by $i$-th charged particle (hereafter
the subscript ``$i$'' means ``$i$-th charged particle''), and $R_{\rm
pe}$ is the photoelectric charging rate. We consider only protons and
electrons as the charged particle.

\subsubsection{Collisional charging rate}

The collisional charging rate by $i$-th charged
particle, $R_i$, is expressed as (e.g., Draine \& Sutin 1987)
\begin{equation}
 R_i = Z_i s_i n_i \int_{0}^{\infty} 
  \sigma_i(a, Z_{\rm d}, Z_i, v_i) v_i f(v_i) dv_i \,,
\end{equation}
where $Z_i$ is the charge in the electron charge unit, $s_i$ is the
sticking coefficient, $n_i$ is the number density, $v_i$ is the
velocity,  $\sigma_i$ is the collisional cross section depending on the
grain radius, $a$, both charges, and the velocity, and $f(v_i)$ is the
velocity distribution function. If the grain and the charged particle
have the charges of the same sign, the kinetic energy of the particle
must exceed the grain electric potential for the collision. Otherwise,
the collisional cross section is zero. We simply assume $s_i$ is always
unity. 

Now, we introduce the dimensionless cross section,
$\tilde{\sigma}_i=\sigma_i/\pi a^2$.
If we neglect the ``image potential'' resulting from the polarization of 
the grain induced by the Coulomb field of the approaching charged
particle (Draine \& Sutin 1987) and we assume the Maxwellian velocity
distribution for the particle and the spherical grains, we obtain
\begin{equation}
 \int_{0}^{\infty} \tilde{\sigma}_i v_i f(v_i) dv_i = 
  \left( \frac{8k_{\rm B}T}{\pi m_i} \right)^{1/2} g(x)\,,
\end{equation}
and
\begin{equation}
 g(x) = \cases{1-x & for $Z_{\rm d}Z_i \leq 0$ \cr
  \exp (-x) & for $Z_{\rm d}Z_i > 0$}\,,
\end{equation}
where $k_{\rm B}$ is the Boltzmann's constant, $T$ is the gas
temperature, $m_i$ is the particle mass, and $x=e^2Z_{\rm d}Z_i/ak_{\rm
B}T$ (Spitzer 1941).

In fact, the ``image potential'' works to enhance the collisional cross
section (Draine \& Sutin 1987).
Although the effect becomes the most important for grains with an around
neutral charge, it quickly declines for highly charged grains.
Indeed, for the charge ratio of $Z_{\rm d}/Z_i < -3$, which is satisfied
in our case as found below, the increment factor for the cross section
by the effect of the ``image potential'' is less than 1.5 (Draine \&
Sutin 1987). Therefore, we neglect the ``image potential'' in this
paper.

\subsubsection{Photoelectric charging rate}

The photoelectric charging rate is given by (e.g., Draine 1978)
\begin{equation}
 R_{\rm pe}=\pi a^2 \int_{0}^{\infty} 
  Q_\nu(a) Y_\nu(a,Z_{\rm d}) \frac{4 \pi J_{\nu}}{h \nu} d\nu\,,
\end{equation}
where $Q_\nu$ is the absorption coefficient of grains at the frequency
$\nu$, $Y_\nu$ is the photoelectric yield, $J_{\nu}$ is the mean
intensity of the incident radiation, and $h$ is the Plank constant.
For $Q_\nu$, we adopt the values of ``graphite'' and ``UV smoothed
astronomical silicate'' by Draine (2003). If the photon energy is
smaller than the threshold energy of the photoelectric emission, e.g.,
the ionization potential or the work function, the yield $Y_\nu=0$.

We adopt a sophisticated model of the photoelectric yield by Weingartner
\& Draine (2001) and Weingartner et al.~(2006) in this paper. The model
of Weingartner \& Draine (2001) (hereafter WD01 model) takes into
account the primary photoelectron emission from the band structure of
grains, a small-size particle effect, and the energy distribution of
the photoelectron. On the other hand, Weingartner et al.~(2006) 
(hereafter W+06 model) add the primary photoelectron emission from 
inner shells of the constituent atoms of grains, the Auger electron
emission, and the secondary electron emission produced by primary
electrons and Auger electrons. The transfer of photons absorbed and
electrons emitted in a grain is also taken into account. The detailed
procedure of the yield calculations is referred to the original papers
of Weingartner \& Draine (2001) and Weingartner et al.~(2006). Fig.~1
shows comparisons between the WD01 and W+06 models. The reduction of the
W+06 yield around 100 eV is due to the effect of the photon/electron
transfer in a grain. The W+06 yield exceeds unity for some cases because
of the Auger and secondary electrons.

\begin{figure}[t]
\centerline{\includegraphics[width=7.0cm]{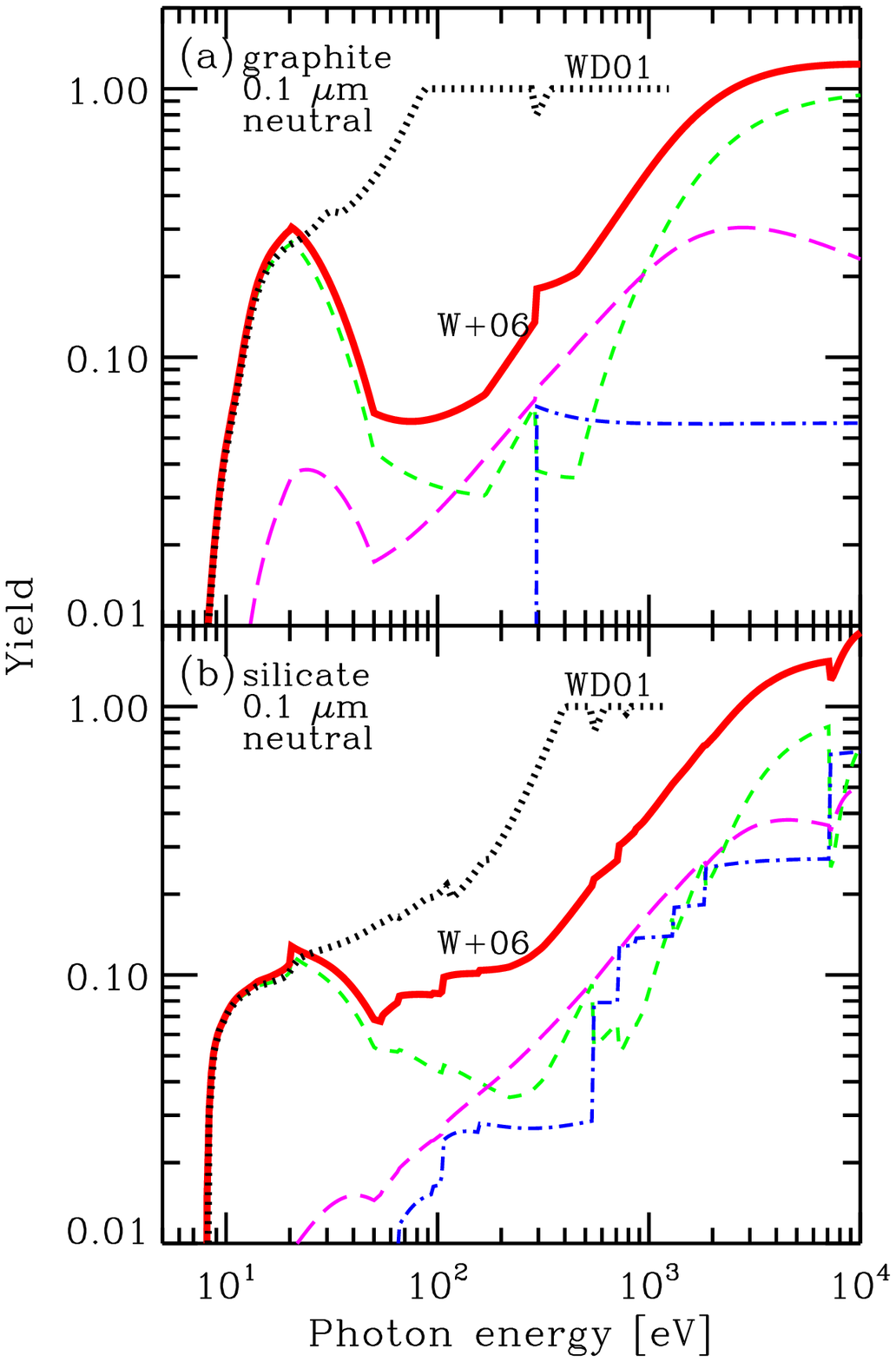}}
\caption{Photoelectric yield models of 0.1 $\mu$m neutral (a)
 graphite and (b) silicate grains. The dotted lines are the WD01
 model (Weingartner \& Draine 2001) and the solid lines are the W+06
 model (Weingartner et al.~2006). The W+06 model consists of three
 processes: the primary photoelectron emission (short dashed line), the
 secondary electron emission (long dashed line), and the Auger electron
 emission (dot-dashed line).}
\end{figure}

We have to note that there is still a large uncertainty of 
photoelectric yield models because of our insufficient understandings of
the nature of small-size particle effect as well as the lack of
experiments. Abbas et al.~(2006) reported measurements of the yield of
individual grains of silica, olivine, and graphite with 0.09--5 $\mu$m
radii for 8--10 eV photons. Their measurements indeed show larger yields
than those of the bulk materials. However, the measurements do not agree
with the yield enhancement factors adopted in WD01 and W+06 models
accounting for the small-size particle effect qualitatively as well as
quantitatively. Clearly, we need more experiments and theoretical
investigations of the photoelectric yield in future.

\subsubsection{Equilibrium charge}

We need to specify the radiation filed incident on grains in the IGM:
the cosmic background radiation. We assume a simple prescription of the
radiation. The intensity of the radiation at the Lyman limit is
estimated from observations of the proximity effect and the Lyman
$\alpha$ forest opacity (e.g., Scott et al. 2000). 
A typical value of the intensity at the
Lyman limit is $J_{\rm L} = 1 \times 10^{-21}$ erg s$^{-1}$ cm$^{-2}$
Hz$^{-1}$ sr$^{-1}$ (e.g., Scott et al.~2000). 
We simply assume a power-law as the
spectral shape: $J_\nu\propto\nu^{-p}$. A typical value of $p$ is unity
(e.g., Haardt \& Madau 1996). With such a radiation filed, the grains in
the IGM are positively charged.

A typical charging time-scale is very short.
For example, the collisional charging rate of electron, 
$R_{\rm e} \sim 5.6\times 10^{-6}$ s$^{-1}$ for 
$n_{\rm e}=10^{-5}$ cm$^{-3}$, $T=10^4$ K,
$a=0.1$ $\mu$m, and $Z_{\rm d}=1700$, which is the equilibrium charge of
graphite or silicate grains for these parameters and 
$J_\nu = 10^{-21} (\nu/\nu_{\rm L})^{-1}$ erg s$^{-1}$
cm$^{-2}$ Hz$^{-1}$ sr$^{-1}$.
Thus, the typical charging time-scale is 
$t \sim 1/R_{\rm e} \sim 6\times10^{-3}$ yr.
Therefore, the grain charge can be in equilibrium. 
We set $dZ_{\rm d}/dt=0$ in equation (1) and obtain the equilibrium
charge of the IGM grains.

\subsection{Heating rates}

\subsubsection{Heating rate per a grain}

The net heating rate per a grain with the radius $a$ is expressed as
(e.g., Weingartner \& Draine 2001) 
\begin{equation}
 \gamma(a)=R_{\rm pe} {\cal E}_{\rm pe}(a) 
  - |R_{\rm e}| {\cal E}_{\rm e}(T)\,,
\end{equation}
where ${\cal E}_{\rm pe}(a)$ is the mean kinetic energy of
photoelectrons from a grain with the radius $a$ and 
${\cal E}_{\rm e}(T)$ is that of electrons colliding with the grain. 
The second term accounts for the cooling by the electron capture. 
If we assume the Maxwellian velocity distribution for the 
electrons, ${\cal E}_{\rm e}(T)=k_{\rm B}T(2+\phi)/(1+\phi)$, where 
$\phi=Z_{\rm d}e^2/ak_{\rm B}T$ (for $Z_{\rm d}>0$; Draine 1978).
We note that ${\cal E}_{\rm e}$ is $\sim1$\% of ${\cal E}_{\rm pe}$ 
in the current setting.

The mean energy of the photoelectrons is given by 
\begin{equation}
 {\cal E}_{\rm pe}(a)= \frac{\pi a^2}{R_{\rm pe}} 
  \int_{0}^{\nu_{\rm max}} Q_\nu(a) YE_\nu(a,Z_{\rm d}) 
  \frac{4 \pi J_{\nu}}{h \nu} d\nu \,,
\end{equation}
and 
\begin{equation}
 YE_\nu(a,Z_{\rm d}) = \sum_k Y^k_\nu(a,Z_{\rm d}) 
  \langle E_{\rm e} \rangle^k_\nu(a,Z_{\rm d})\,,
\end{equation}
where $Y^k_\nu$ is the photoelectric yield of $k$-th emission process,
e.g., primary electrons from the band structure, Auger electron, etc., 
and $\langle E_{\rm e} \rangle^k_\nu$ is the mean energy of electrons
emitted by $k$-th process with the absorbed photon energy $h\nu$.
The estimation of $\langle E_{\rm e} \rangle^k_\nu$ is based on the
assumed energy distribution of the electrons. Following Weingartner et
al.~(2006), we adopt a parabolic function for the primary and the auger 
electrons and a function introduced by Draine \& Salpeter (1979) for the
secondary electrons, which were derived to fit some experimental results.

In the IGM, the grains are positively charged. Then, the proton
collisional charging rate is negligible. Thus, the photoelectric
charging rate balances with the electron collisional charging rate: 
$R_{\rm pe}+R_{\rm e}=0$. In this case, equation (6) is reduced to 
\begin{eqnarray}
 \gamma(a) & = &|R_{\rm e}|({\cal E}_{\rm pe} - {\cal E}_{\rm e}) \cr
  & \approx &\pi a^2 n_{\rm e} 
  \left(\frac{8k_{\rm B}T}{\pi m_{\rm e}} \right)^{1/2} 
  \left(\frac{eV_{\rm d}}{k_{\rm B}T}\right){\cal E}_{\rm pe}\,, 
\end{eqnarray}
where we have used equations (2--4) for $R_{\rm e}$ and 
$V_{\rm d}=Z_{\rm d}e/a$ is the grain electric potential 
($eV_{\rm d}/k_{\rm B}T\gg1$ and ${\cal E}_{\rm pe}\gg{\cal E}_{\rm e}$
for the IGM). As found later in Fig.~2, the
electric potential depends on the grain size weakly in the W+06 yield
case. We confirmed that the mean energy of photoelectrons also 
depends on the grain size weakly. As a result, the heating rate per a
grain is roughly proportional to the square of the size, which is shown
later in Fig.~3.

\subsubsection{Total photoelectric heating rate}

To estimate the total photoelectric heating rate per unit volume, 
we need to specify the amount and the size distribution of dust grains.
A power law type distribution for grain size is familiar in the
interstellar medium of the Milky Way since classical work by Mathis,
Rumpl, \&  Nordsieck (1977; hereafter MRN).
Power law is expected to be achieved by resulting from
coagulation, shattering, and sputtering processes (e.g., Jones, Tielens,
\& Hollenbach 1996).
Here we express the power law distribution as $n(a)=Aa^{-q}$,
where $n(a)da$ is the number density of grains with the radius
between $a$ and $a+da$. For the MRN distribution, $q=3.5$ 
(see Table 1). The normalization $A$ is determined from the
total dust mass density 
$\rho_{\rm d}=\int_{a_{\rm min}}^{a_{\rm max}}m(a)n(a)da$, 
where $m(a)=(4\pi/3)\varrho a^3$ is the mass of grains with the radius
$a$, $\varrho$ $(\simeq 3~{\rm g\,cm^{-3}})$ is the grain material
density, $a_{\rm min}$ and $a_{\rm max}$ are the minimum and maximum
radius, respectively. The dust mass density $\rho_d$ is given by 
$\rho_{\rm d}=m_{\rm p}n_{\rm H}{\cal D}$, where $m_{\rm p}$ is the
proton mass, $n_{\rm H}$ is the hydrogen number density, and 
${\cal D}$ is the dust-to-gas mass ratio. We assume ${\cal D}=10^{-4}$, 
which is about two orders of magnitude smaller than that in the Milky
Way's ISM. Then, the total photoelectric heating rate is 
\begin{equation}
 \Gamma_{\rm pe} = \int_{a_{\rm min}}^{a_{\rm max}}
  \gamma(a) n(a) da\,.
\end{equation}

Let us consider a typical size for the total heating rate. Using the
grain number density $n_{\rm d}=\int_{a_{\rm min}}^{a_{\rm max}}n(a)da$, 
we can define a mean heating rate per a grain as $\langle \gamma \rangle
\equiv \int_{a_{\rm min}}^{a_{\rm max}}\gamma(a)n(a)da/n_{\rm d}$ and a
mean mass per a grain as $\langle m_{\rm d} \rangle \equiv \rho_{\rm d}
/n_{\rm d} = \int_{a_{\rm min}}^{a_{\rm max}}m(a)n(a)da/n_{\rm d}$. 
Then, the total heating rate is reduced to $\Gamma_{\rm pe} = 
\langle \gamma \rangle n_{\rm d} = \langle \gamma \rangle \rho_{\rm d} 
/ \langle m_{\rm d} \rangle$. The heating rate per a grain can be
approximated to $\gamma(a)\approx\gamma_0a^2$ as seen in \S2.2.1 (see
also Fig.~3 and \S3.1.2), where $\gamma_0$ is a normalization. The grain
mass is $m(a)=(4\pi/3)\varrho a^3$. Then, we obtain 
\begin{equation}
 \Gamma_{\rm pe} \approx \frac{3 \rho_{\rm d} \gamma_0}
  {4\pi \varrho \langle a \rangle}\,,
\end{equation}
where a typical size $\langle a \rangle$ is given by 
\begin{equation}
 \langle a \rangle = \frac{\int_{a_{\rm min}}^{a_{\rm max}} a^3 n(a)da}
  {\int_{a_{\rm min}}^{a_{\rm max}} a^2 n(a)da}\,.
\end{equation}
Note that a larger typical size results in a smaller total heating rate
because of a smaller number density of grains for a fixed dust mass.

\subsubsection{Photoionization heating rates}

For comparison with the photoelectric heating rate by grains, we
estimate the photoionization heating rates of hydrogen and helium.
The net HI photoionization heating rate is 
\begin{equation}
 \Gamma_{\rm pi}^{\rm HI}=n_{\rm HI} R^{\rm HI}_{\rm pi} 
  {\cal E}^{\rm HI}_{\rm pi} - n_{\rm HII} R^{\rm HI}_{\rm re} 
  {\cal E}_{\rm gas}\,,
\end{equation}
where $R^{\rm HI}_{\rm pi}=\int_{\nu^{\rm HI}_{\rm L}}^{\infty} 
\sigma^{\rm HI}_\nu 4 \pi J_\nu/h\nu d\nu$ is the HI photoionization
rate, $R^{\rm HI}_{\rm re}=n_{\rm e} \alpha^{\rm HI}_{\rm A}(T)$ is the
HI recombination rate, ${\cal E}^{\rm HI}_{\rm pi} = 
(1/ R^{\rm HI}_{\rm pi})\int_{\nu^{\rm HI}_{\rm L}}^{\infty} 
\sigma^{\rm HI}_\nu 4 \pi J_\nu/h\nu (h\nu-h\nu_{\rm L}^{\rm HI})d\nu$ 
is the mean kinetic energy of the HI photoionized electrons, 
$\sigma^{\rm HI}_\nu$ is the HI photoionization cross section, 
$\nu^{\rm HI}_{\rm L}$ is the HI Lyman limit frequency, $n_{\rm HI}$, 
$n_{\rm HII}$, and $n_{\rm e}$ are the neutral hydrogen, ionized
hydrogen, and electron number densities, respectively, 
$\alpha_{\rm A}^{\rm HI}(T)$ is the Case A HI recombination coefficient
for the gas temperature $T$ (Osterbrock \& Ferland 2006), and 
${\cal E}_{\rm gas}$ is the mean kinetic energy lost from the gas per
one recombination. If we assume that $J_\nu \propto \nu^{-p}$ 
and $\sigma^{\rm HI}_\nu \propto \nu^{-3}$, we obtain 
${\cal E}^{\rm HI}_{\rm pi} = h\nu^{\rm HI}_{\rm L}/(p+2)$.
If we take into account the gas cooling by free-free emission, 
${\cal E}_{\rm gas}\approx k_{\rm B}T$ for the Case A and $T=10^4$ K 
(Osterbrock \& Ferland 2006). If we assume the ionization equilibrium, 
$n_{\rm HI}R^{\rm HI}_{\rm pi}=n_{\rm HII}R^{\rm HI}_{\rm re}$, 
we obtain 
\begin{equation}
 \Gamma^{\rm HI}_{\rm pi} = n_{\rm H}^2 \alpha^{\rm HI}_{\rm A}(T)
  ({\cal E}^{\rm HI}_{\rm pi} - {\cal E}_{\rm gas}), 
\end{equation}
where we have assumed $n_{\rm HII}=n_{\rm e}=n_{\rm H}$ with $n_{\rm H}$
being the hydrogen number density, that is, the neutral fraction is
assumed to be very small. The net HeII photoionization heating rate is
likewise 
\begin{equation}
 \Gamma^{\rm HeII}_{\rm pi} = n_{\rm He} n_{\rm H} 
  \alpha^{\rm HeII}_{\rm A}(T)
  ({\cal E}^{\rm HeII}_{\rm pi} - {\cal E}_{\rm gas}), 
\end{equation}
$n_{\rm He}$ is the helium number density, $\alpha^{\rm HeII}_{\rm A}(T)$ 
is the HeII recombination rate, and ${\cal E}^{\rm HeII}_{\rm pi}$ is
the mean kinetic energy of the HeII photoionized electrons.
We assume $n_{\rm He}/n_{\rm H}=0.1$.

\section{Results}

\subsection{Comparison between the two yield models}

We compare the grain charge and heating rates with the WD01 and W+06
models quantitatively in the IGM environment. Weingartner et al.~(2006)
showed the grain charges in QSO environments which are similar
situation with a similar radiation field in this paper. However, they
did not show the heating rates in the environment.

\subsubsection{Electric potential}

\begin{figure}[t]
\centerline{\includegraphics[width=7.0cm]{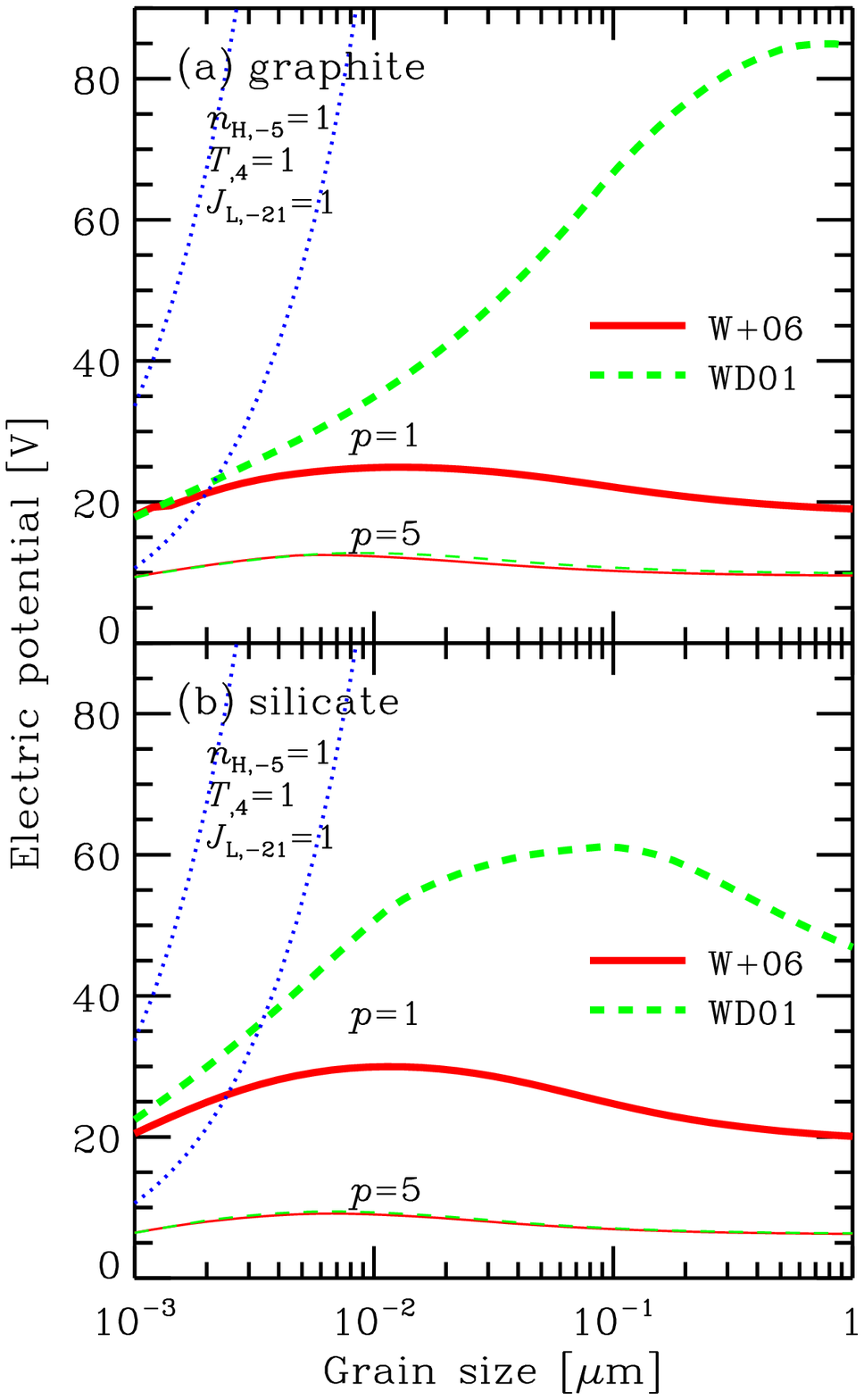}}
\caption{Equilibrium electric potential as a function of grain size: (a)
 graphite and (b) silicate. The solid lines are the W+06
 model and the dashed lines are the WD01 model. The thick lines are the
 case with the spectral index of the radiation field $p=1$ and the thin
 lines are the case with $p=5$. Other assumed quantities are noted in the
 panels as the hydrogen density $n_{\rm H,-5}=n_{\rm H}/10^{-5}~{\rm cm^{-3}}$, 
 the gas temperature $T_{,4}=T/10^4~{\rm K}$, and the radiation intensity 
 $J_{\rm L,-21}=J_{\rm L}/10^{-21}~{\rm erg~s^{-1}~cm^{-2}~Hz^{-1}~sr^{-1}}$.
 The dotted lines show the critical electric potential where the grain
 destruction occurs by the Coulomb explosion; the upper lines are the
 case with the tensile strength of $10^{11}$ dyn cm$^{-2}$ and the lower
 lines are the case with $10^{10}$ dyn cm$^{-2}$.}
\end{figure}

In Fig.~2, we compare the electric potentials of the W+06 model (solid
lines) with those of the WD01 model (dashed lines). We showed two cases
of the spectrum of the radiation field; one has a hard spectrum as a
background radiation dominated by QSOs which is the case with the
spectral index $p=1$, and the other has a soft spectrum with $p=5$ for a
comparison. Other assumed quantities are appropriate for the IGM at the
redshift $z\sim3$ and shown in the panels. 
The radiation fields assumed here correspond to the ionization parameter 
$U\equiv n_{\rm ion}/n_{\rm H}$, which is the number density ratio of
ionizing photons and hydrogen nucleus, of 6.3 for $p=1$ and of 1.3 for
$p=5$. Weingartner et al.~(2006) showed electric potentials in their
Figs.~6 and 7 with $U=0.1$--100. We find that our calculations are
quantitatively well matched with theirs.

We find in Fig.~2 that for the hard spectrum case, the grain electric
potentials with the W+06 yield model are much smaller than those with
the WD01 model, especially for larger grain sizes. On the other hand,
for the soft spectrum case, the difference is very small, less than
4\%. This is because the main difference between the W+06 yield and the
WD01 yield is found in the primary photoelectron yield at $\sim100$ eV
due to the photon/electron transfer in a grain as shown in Fig.~1. In
the soft spectrum case, since there are not many photons around the
energy, we do not find a significant difference between the two yield
models. For smaller grain sizes, the yield reduction by the
photon/electron transfer is small as found in Figs.~4 and 5 of
Weingartner et al.~(2006). Thus, we do not find a significant difference
in the electric potentials for smaller grain sizes in Fig.~2, either.

The electrostatic stress on a grain may cause the grain destruction by
the Coulomb explosion (e.g., Draine \& Salpeter 1979). The critical
electric potential is $V_{\rm max} = 1063~{\rm V}~(S_{\rm d}/10^{10}
~{\rm dyn~cm^{-2}})^{1/2}(a/0.1~{\rm \mu m})$, 
where $S_{\rm d}$ is the tensile strength of grains, which is very
uncertain. Perfect crystal structure may have $S_{\rm d}\sim10^{11}$
dyn cm$^{-2}$ (Draine \& Salpeter 1979), but imperfections would reduce
the strength as $S_{\rm d}\sim10^{10}$ dyn cm$^{-2}$ (Fruchter et
al.~2001). Following Weingartner et al.~(2006), we show two cases of the
critical potential with $S_{\rm d}\sim10^{10}$ and $10^{11}$ dyn
cm$^{-2}$ in Fig.~2 as the dotted lines. The critical potential by the
ion field emission is similar to the case with $S_{\rm d}\sim10^{11}$
dyn cm$^{-2}$ (Draine \& Salpeter 1979). We find that grains smaller
than 20--30 \AA\ in the hard radiation field may be destroyed by the
Coulomb explosion. Then, there may be no very small grains in the IGM.

\subsubsection{Photoelectric heating rate}

\begin{figure}[t]
\centerline{\includegraphics[width=7.0cm]{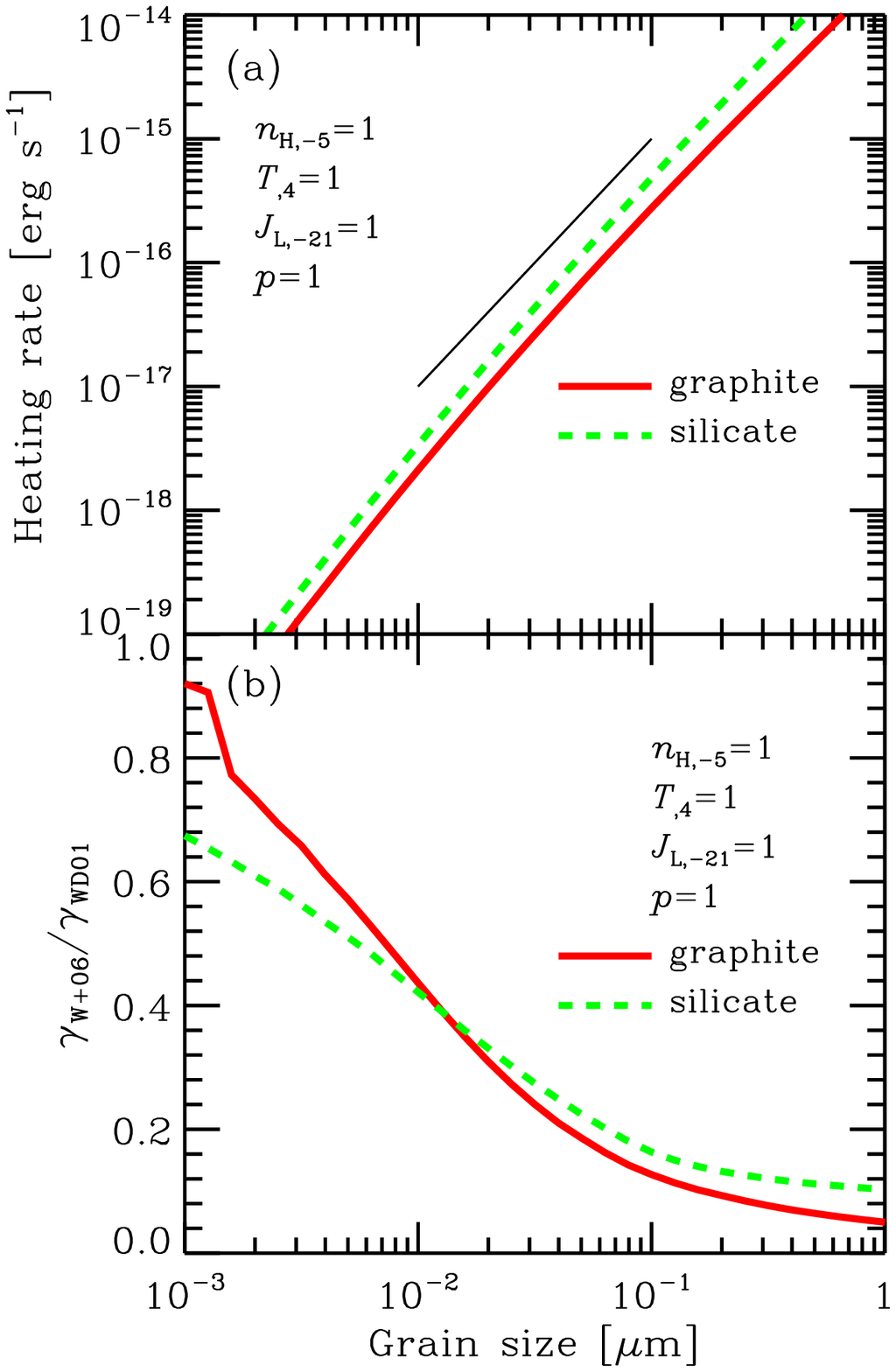}}
\caption{Photoelectric heating rate per a grain as a function of grain
 size: (a) heating rate with the W+06 yield model and (b) ratio of the
 heating rate with the W+06 yield model to that with the WD01 yield
 model. The solid lines are the graphite case and the dashed lines
 are the silicate case. The assumed quantities are noted in the
 panels as the hydrogen density $n_{\rm H,-5}=n_{\rm H}/10^{-5}~{\rm cm^{-3}}$, 
 the gas temperature $T_{,4}=T/10^4~{\rm K}$, the radiation intensity 
 $J_{\rm L,-21}=J_{\rm L}/10^{-21}~{\rm erg~s^{-1}~cm^{-2}~Hz^{-1}~sr^{-1}}$, 
 and the spectral index of the radiation field $p=1$. The thin solid
 line in the panel (a) shows the slope proportional to the square of size.}
\end{figure}

Fig.~3 shows the photoelectric heating rate per a grain in a typical
$z\sim3$ IGM environment with a hard radiation; graphite
grains are shown by solid lines and silicate grains are shown by the
dashed lines. In the panel (a), we show the absolute value of the
heating rate for the W+06 yield model. As expected in equation (9), the
heating rate is nicely proportional to $a^2$, square of the size. 
However, the slope becomes gradually steep for a small ($<100$ \AA)
grain size. 

In the panel (b), we compare the two heating rates with the W+06 model
and the WD01 model. We find that the heating rate with the W+06 model is
much smaller than that with the WD01 model: a factor of 10 smaller
for the largest grain size. This is because (1) reduction of the grain
electric potential and (2) reduction of the mean energy of the
photoelectron in the W+06 model. As found in equation (9), the heating
rate per a grain is proportional to the product of the potential and
the mean photoelectron energy. As seen in Fig.~2, the W+06 model expects
smaller potential because of the reduction of the yield at $\sim100$
eV. The yield reduction also causes the reduction of the mean energy of
the photoelectron as expected in equation (8). Therefore, we have up to
about a factor of 10 reduction of the heating rate with the W+06 model.

\begin{figure}[t]
\centerline{\includegraphics[width=7.0cm]{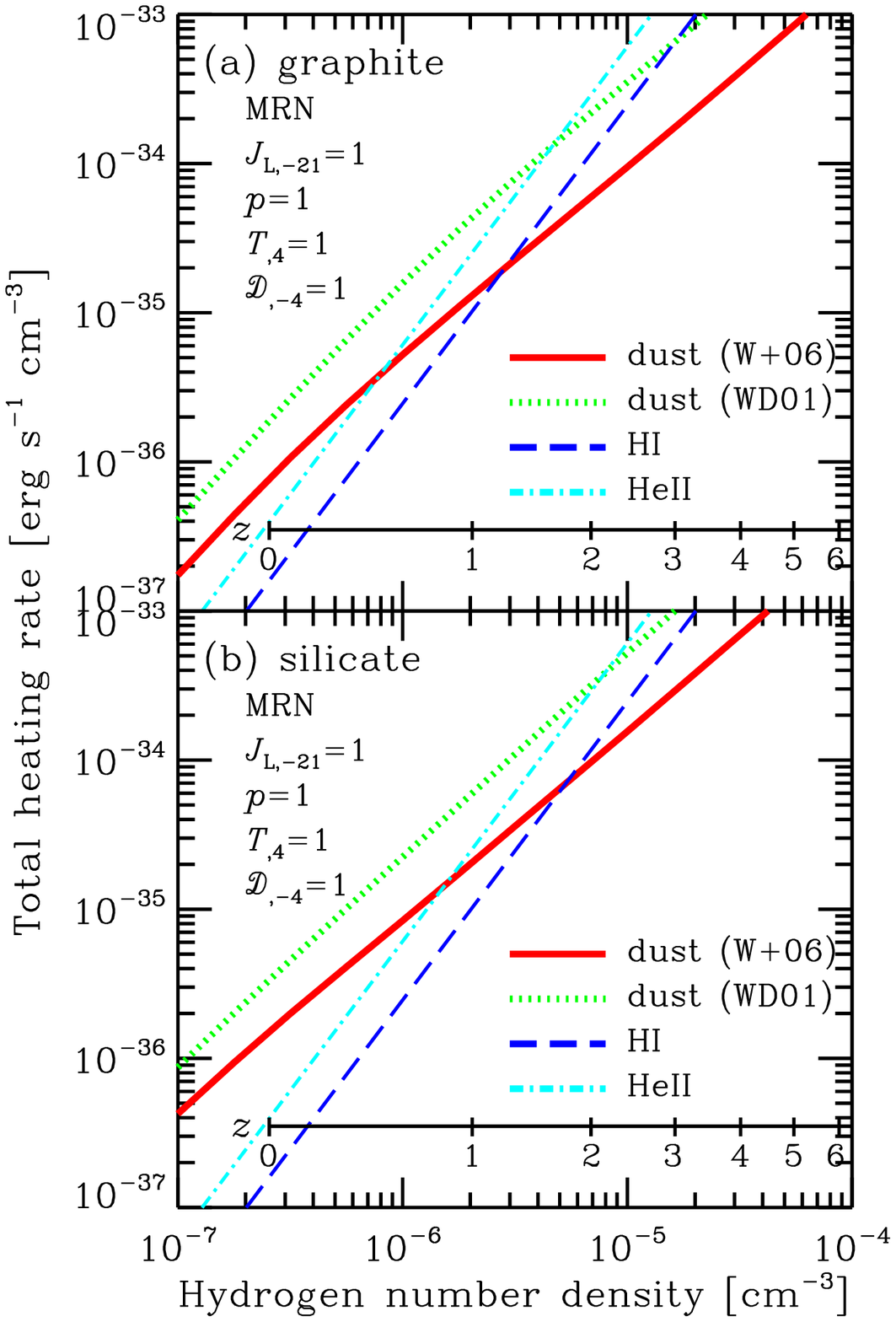}}
\caption{Photoelectric heating rates as a function of hydrogen number
 density: (a) graphite and (b) silicate. The solid lines
 are the W+06 model and the dotted lines are the WD01 model. The assumed
 grain size distribution is the so-called MRN distribution (Mathis,
 Rumpl, \& Nordsieck 1977). Other assumed quantities are noted in the
 panels as the gas temperature $T_{,4}=T/10^4~{\rm K}$, the radiation
 intensity $J_{\rm L,-21}=J_{\rm L}/10^{-21}~{\rm
 erg~s^{-1}~cm^{-2}~Hz^{-1}~sr^{-1}}$, the spectral index of the
 radiation field $p$, and the dust-to-gas mass ratio 
 ${\cal D}_{,-4}={\cal D}/10^{-4}$. The dashed lines are the HI
 photoionization heating rate and the dot-dashed lines are the HeII
 photoionization heating rate with assuming the ionization
 equilibrium. We also show the redshift at which the number density on
 the horizontal axis corresponds to the mean density of the Universe.}
\end{figure}

Fig.~4 shows a comparison of the total heating rates by the W+06 model
(solid lines) and by the WD01 model (dotted lines). The horizontal axis
is the assumed hydrogen number density. We also show the redshift at
which the number density on the horizontal axis corresponds to the mean
density of the Universe. We have assumed the MRN grain size distribution
(see Table 1). We find that the total heating rate with the W+06 yield
is a factor of 2--4 smaller than that with the WD01 yield.

For a comparison, we also show the HI and HeII photoionization heating
rates in Fig.~4. We have assumed the ionization equilibrium for them.
When we assume the dust-to-gas ratio in the IGM is 1\% of that in the
Milky Way (i.e. ${\cal D}=10^{-4}$), the dust photoelectric heating
dominates the HI and HeII photoionization heatings if the hydrogen
number density is less than $10^{-6}$--$10^{-5}$ cm$^{-3}$ which
correspond to the redshift $z\sim1$--2. We note that the dust heating is
the most important mechanism in the IGM at $z=0$ even with the W+06
yield model if the IGM has dust with 1\% dust-to-gas ratio of the Milky
Way and with the MRN size distribution.

\subsection{Effect of the grain size distribution}

The size distribution of the intergalactic dust grains should be
important for the photoelectric heating rate via the typical size
defined by equation (12). However, it is quite uncertain. Thus, we
examine several possibilities of the size distribution in this
section. Table 1 is a summary of the size distribution considered here.

\begin{table}[t]
 \renewcommand{\arraystretch}{1.2}
 \vspace{-.3cm}
 \caption{Possible size distributions of the intergalactic dust.}
 \vspace{-.1cm}
 \begin{center}
  \begin{tabular}{lc} \hline
   MRN & Mathis, Rumpl, \& Nordsieck (1977) \\ \hline
   \multicolumn{2}{l}{single power law$^a$} \\
   $q$ & $3.5$ \\
   $a_{\rm min}$ & 50 \AA \\
   $a_{\rm max}$ & 0.25 $\mu$m \\ 
   $\langle a \rangle$ & 350 \AA \\ \hline
   BF05 & Bianchi \& Ferrara (2005) \\ \hline
   \multicolumn{2}{l}{single power law$^a$} \\
   $q$ & $3.5$ \\
   $a_{\rm min}$ & 0.1 $\mu$m \\
   $a_{\rm max}$ & 0.25 $\mu$m \\ 
   $\langle a \rangle$ & 0.16 $\mu$m \\ \hline
   N03 & Nozawa et al.~(2003) \\ \hline
   \multicolumn{2}{l}{double power law$^b$} \\
   $q_1$ ($a\leq a_{\rm c}$) & $2.5$  \\
   $q_2$ ($a>a_{\rm c}$) & $3.5$ \\
   $a_{\rm min}$ & 2 \AA \\
   $a_{\rm max}$ & 0.3 $\mu$m \\ 
   $a_{\rm c}$ & 0.01 $\mu$m \\ 
   $\langle a \rangle$ & 290 \AA \\ \hline
   N07 & Nozawa et al.~(2007) \\ \hline
   \multicolumn{2}{l}{double power law$^b$} \\
   $q_1$ ($a\leq a_{\rm c}$) & $1.0$  \\
   $q_2$ ($a>a_{\rm c}$) & $2.5$ \\
   $a_{\rm min}$ & 10 \AA \\
   $a_{\rm max}$ & 0.3 $\mu$m \\ 
   $a_{\rm c}$ & 0.01 $\mu$m \\ 
   $\langle a \rangle$ & 0.12 $\mu$m \\ \hline
   SG & --- \\ \hline
   \multicolumn{2}{l}{single power law$^a$} \\
   $q$ & $3.5$ \\
   $a_{\rm min}$ & 50 \AA \\
   $a_{\rm max}$ & 0.025 $\mu$m \\ 
   $\langle a \rangle$ & 110 \AA \\ \hline
  \end{tabular}
 \end{center}
 $^a$ The grain size distribution $n(a) \propto a^{-q}$.\\
 $^b$ The grain size distribution $n(a) \propto a^{-q_1}$ for 
 $a\leq a_{\rm c}$ and $\propto a^{-q_2}$ for $a> a_{\rm c}$.
\end{table}

The grain size distribution in the Milky Way has been approximated to be
a power-law since Mathis et al.~(1977) suggested as $n(a)\propto a^{-q}$
with $q=3.5$. This MRN distribution is a reference case and is already
adopted in Fig.~4. During the grain transport from galaxies to the IGM,
there may be size filtering mechanisms. For example, Ferrara et
al.~(1991) showed that sputtering in the hot gas filling the galactic
halo efficiently destroys grains smaller than $\sim0.1$ $\mu$m. Bianchi
\& Ferrara (2005) also showed that only grains larger than $\sim0.1$
$\mu$m reach a significant distance (a few $\times$ 100 kpc) from the
parent galaxies by calculating the grain ejection by the radiation
pressure and the grain destruction by the sputtering simultaneously.
Here, we consider a simple size distribution of the MRN with $\ge0.1$
$\mu$m grains as the BF05 model.

In the early Universe, the dominant source of dust grains is 
different from that in the current Milky Way. Although asymptotic giant
branch stars are considered to be the main dust source in the Milky Way
(e.g., Dwek 1998), there is not enough time for stars to evolve to the
phase in the early Universe at the redshift $z>6$. However, a plenty of
dust is found in QSOs at $z>6$ (Bertoldi et al.~2003). SNe are the
candidate of the dust source in the early Universe (e.g., Nozawa et
al.~2003) and the observed extinction curve of dust associated with the
QSO is compatible with those expected from the grains produced by SNe 
(Maiolino et al.~2004, Hirashita et al.~2005,2008). Thus, we consider
the size distribution expected from the SNe dust production model by
Nozawa et al.~(2003) as the N03 model. In addition, we adopt the size
distribution expected by Nozawa et al.~(2007), who explored the effect
of the dust destruction by the reverse shock in the SN remnant, as the
N07 model.

Finally, we adopt a hypothetical size distribution consisting of only
small grains as a comparison case; the MRN distribution with the maximum
size of 250 \AA\ as the SG (small grain) model.

\begin{figure}[t]
\centerline{\includegraphics[width=7.0cm]{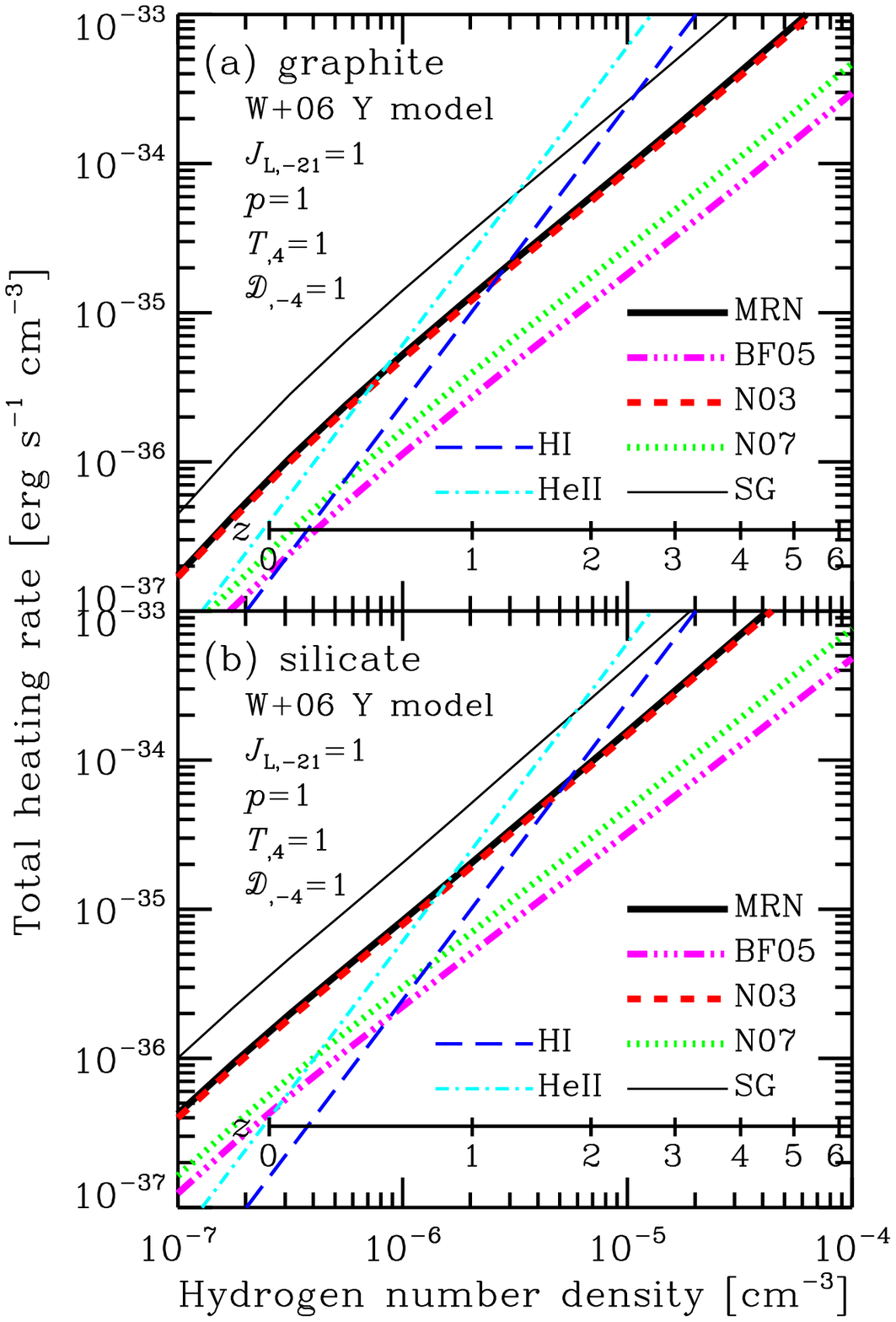}}
\caption{Same as Fig.~4 but for various size distribution functions with
 the W+06 yield model: (a) graphite and (b) silicate. 
 The thick solid lines are the MRN case. The short-dashed lines
 are the size distribution expected from the grain formation model in
 supernova ejecta by Nozawa et al.~(2003). The dotted lines are the size
 distribution expected after the grain destruction by the reverse shock
 in the supernova remnant by Nozawa et al.~(2007). The triple-dot-dashed
 lines are the MRN but only size larger than 0.1 $\mu$m because of a
 filtering effect in the transfer of grains from galaxies to the IGM
 suggested by Bianchi \& Ferrara (2005). The thin solid lines are the
 MRN but only size smaller than 250 \AA\ as a comparison. The dashed and
 dot-dashed lines are HI and HeII heating rates.}
\end{figure}

Fig.~5 shows a comparison of total heating rates with the five size
distributions considered here. All the cases are assumed the W+06 yield
model and physical conditions appropriate for the IGM. The case of the
BF05 model (triple-dot-dashed line) is a factor of $\simeq5$ smaller
than that of the MRN model (thick solid line). This reduction factor is
simply accounted for by the ratio of the typical sizes of the two
models: 0.16 $\mu$m for the BF05 model and 350 \AA\ for the MRN
model (see Table 1). The same thing is true for the N07 model (dotted
line) and the SG model (thin solid line). The result of the N03 model
(dashed line) coincides with that of the MRN model because their typical
sizes are similar. In any case, we have a smaller number of grains for a
larger typical size if the total dust mass is fixed. Then, the heating
rate is reduced. We note that the dust photoelectric heating is still
dominant or important mechanism relative to the HI and HeII
photoionization heatings in the $z=0$ IGM even with the BF05 model if
the dust-to-gas ratio in the IGM is 1\% of that in the Milky Way.

\subsection{A simple formula of the dust photoelectric heating rate}

\begin{figure}[t]
\centerline{\includegraphics[width=7.0cm]{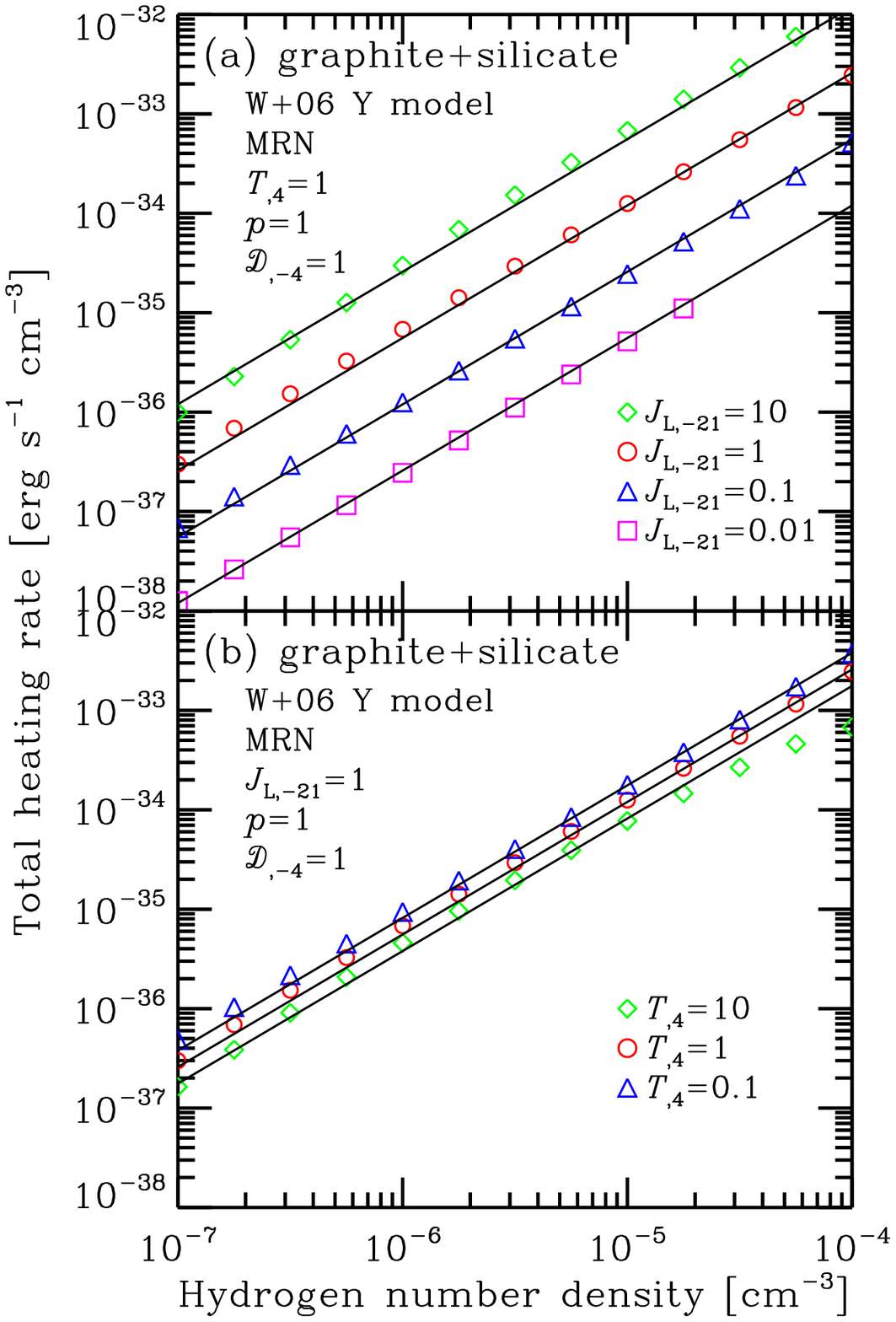}}
\caption{Same as Fig.~4 but for various settings. The photoelectric
 yield model is the W+06 model. We assume that the dust consists of a
 mixture of graphite and silicate (50\% each in mass) with the MRN size
 distribution. (a) Different intensities at the Lyman limit of the
 radiation field: $J_{\rm L}/10^{-21}$ erg s$^{-1}$ cm$^{-2}$ Hz$^{-1}$
 sr$^{-1}$ $=10$ (diamonds), 1 (circles), 0.1 (triangles), and 0.01
 (squares). (b) Different temperatures of the gas: $T/10^4$ K $=10$
 (diamonds), 1 (circles), and 0.1 (triangles). Other assumed quantities
 are noted in the panels. See the caption of Fig.~4 for the notation.
 The solid lines are the simple formula shown in eq.~(16).}
\end{figure}

Fig.~6 shows the effect of different settings of the calculation on the
dust photoelectric heating rate: (a) various intensities of the
background radiation and (b) various temperatures of the gas. The W+06
yield model and the MRN size distribution are assumed. We also assume
that the dust consists of a mixture of graphite and silicate with the
mass ratio of 1:1. The spectral index of the background radiation is
always set unity. In the weakest intensity case (squares in the panel
[a]), the equilibrium charges for smallest grains ($<0.01$ $\mu$m) are
less than 3 in the electric charge unit for the hydrogen density 
$n_{\rm H}>2\times10^{-5}$ cm$^{-3}$. In these cases, the effect of the
``image potential'' (Draine \& Sutin 1987) is not negligible, then, the
current calculations are no longer valid. We note that all the cases
shown in Fig.~6 have an equilibrium charge much larger than 3 for all
grains in the size distribution.

The resultant heating rates are well expressed as 
\begin{eqnarray}
 \Gamma_{\rm pe} & = & 1.2\times10^{-34}~{\rm erg~s^{-1}~cm^{-3}} \cr 
  & \times & \left(\frac{\cal D}{10^{-4}}\right) 
  \left(\frac{n_{\rm H}}{10^{-5}~{\rm cm^{-3}}}\right)^{4/3}
  \left(\frac{T}{10^4~{\rm K}}\right)^{-1/6} \cr
  & \times & \left(\frac{J_{\rm L}}{10^{-21}~
   {\rm erg~s^{-1}~cm^{-2}~Hz^{-1}~sr^{-1}}}\right)^{2/3}\,,
\end{eqnarray}
which is shown in Fig.~6 as solid lines. The indices in this formula can
be derived analytically following Inoue \& Kamaya (2004). From eqs.~(A4)
and (A7) in Inoue \& Kamaya (2004), we find $\Gamma_{\rm pe}\propto 
J_{\rm L}^{2/(p+\beta+1)} n_{\rm H}^{2-2/(p+\beta+1)} 
T^{3/2-(2p+2\beta+1)/(p+\beta+1)}$, where $\beta$ is the emissivity (or
absorption) index of the dust: $Q_\nu \propto \nu^{-\beta}$. Here, we
have $p=1$ and $\beta\approx1$, then, we obtain the indices in eq.~(16).

The deviation of the heating rates from the formula for $T=10^5$ K and 
$n_{\rm H}>2\times10^{-5}$ cm$^{-3}$ is due to the relative significance
of the cooling by the electron capture (see eq.~[6]). Indeed, we find
that the mean energy of photoelectrons from small ($<0.01$ $\mu$m)
graphite grains is smaller than the mean kinetic energy of the $10^5$ K
gas in the case of $n_{\rm H}=10^{-4}$ cm$^{-3}$, then, the net heating
rate per such a graphite grain is negative. Thus, we have a reduction of
the total heating rate for $T=10^5$ K and $n_{\rm H}>2\times10^{-5}$ 
cm$^{-3}$ as found in Fig.~6 (b) although the heating rate by silicate
grains is still positive. 

The validity of the formula presented in equation~(16) is ensured for 
$n_{\rm H}=10^{-7}$--$10^{-4}$ cm$^{-3}$, $J_{\rm L}=10^{-23}$--$10^{-20}$ 
erg s$^{-1}$ cm$^{-2}$ Hz$^{-1}$ sr$^{-1}$, and $T=10^3$--$10^5$ K
within a uncertainty of 30\%, except for $n_{\rm H}>2\times10^{-5}$ 
cm$^{-3}$ with $J_{\rm L}=10^{-23}$ erg s$^{-1}$ cm$^{-2}$ Hz$^{-1}$
sr$^{-1}$ or $T=10^5$ K. Note that there may be much larger uncertainty
in the photoelectric yield model. If one likes another size distribution
rather than the standard MRN, for example the BF05 model discussed in
\S3.3, the heating rate might be scaled by a factor found in Fig.~5 or
the ratio of the typical sizes in Table~1.

\section{Discussions}

\subsection{Amount of the intergalactic dust}

Inoue \& Kamaya (2003,2004) discussed the effect of the photoelectric
heating by the intergalactic dust on the thermal history of the IGM, and
then, obtained an upper limit of the intergalactic dust amount. However,
we have already seen that the W+06 yield model results in a factor of
2--4 reduction of the photoelectric heating rate relative to the WD01
model which was adopted in Inoue \& Kamaya (2003,2004). We can conclude
that the upper limits obtained from the IGM thermal history are raised
by a few factor.
Even in this case, the final limit obtained by Inoue \& Kamaya (2004),
which is that the intergalactic dust mass should be less than 10\% of
the metal mass produced in galaxies, is not affected because the limit
was obtained mainly from the reddening measurements of SNe Ia at
$z=0.5$, especially for $\sim0.1$ $\mu$m size grains.

\subsection{Can grains cause an inverted temperature--density relation
  in the IGM?}

Bolton et al.~(2008) recently suggest an inverted temperature--density
relation in the low density IGM at $z=2$--3. Previously the temperature
in the low density IGM is thought to be proportional to the density
positively (e.g., Hui \& Gnedin 1997). However, Bolton et al.~(2008)
examined carefully the probability distribution function (PDF) of the
flux in QSOs' spectra through the Lyman $\alpha$ forest in the IGM, and
found that the observed PDF is explained better by the negatively
proportional temperature--density relation; lower density IGM is
hotter. This needs a more efficient heating source for lower density
IGM. Bolton et al.~(2008) suggested a radiation transfer effect
(e.g., Abel \& Haehnelt 1999) for the mechanism.

The intergalactic dust may contribute to the heating in the low density
IGM. As shown in Figs.~4 and 5, the importance of the dust photoelectric
heating increases in lower density medium which is plausible for the
inverted temperature--density relation. For example, we expect a factor
of $\sim2$ larger heating rate by dust than HeII photoionization heating
in a medium with 1/10 of the mean density at $z=2$ for the MRN size
distribution and 1\% dust-to-gas ratio of the Milky Way.
Thus, the dust photoelectric heating may cause the inverted
temperature--density relation observed in the low density IGM at
$z=2$--3. This point should be examined further by implementing the dust
heating in a cosmological hydrodynamics simulation. For this, the
formula presented in equation~(16) will be useful.

\subsection{Photoelectric effect before the cosmic reionization}

Finally, we examine if the dust photoelectric heating is efficient in
the IGM before the cosmic reionization. Because of the prominent
Gunn--Peterson trough in QSOs' spectra (e.g., Fan et al.~2006), the
cosmic reionization epoch should be at $z>6$. Here we consider the IGM
at $z\sim10$.

Before the reionization, the ionizing background radiation does not
exist although a nonionizing UV background can be established by
primordial galaxies or active blackhole--accretion disk systems. 
An X-ray background radiation may also exist (e.g., Venkatesan et 
al.~2001). We consider two cases; one is the case with only
nonionizing UV background radiation and the other is the case with
additional X-ray background radiation. For simplicity, we assume the
background radiation to be a power-law with the spectral index $p=1$ and
the intensity at the Lyman limit $J_{\rm L}=1\times10^{-21}$ erg
s$^{-1}$ cm$^{-2}$ Hz$^{-1}$ sr$^{-1}$. However, we assume no intensity
between $E_{\rm UV}^{\rm max}=13.6$ eV and $E_{\rm X}^{\rm min}=300$ eV.
Thus, in the nonionizing UV only case, we have the background radiation
only below $E_{\rm UV}^{\rm max}=13.6$ eV. In the case with the X-ray
background, we have radiation below $E_{\rm UV}^{\rm max}=13.6$ eV and
above $E_{\rm X}^{\rm min}=300$ eV. The dust-to-gas ratio in the IGM at
$z\sim10$ is of course unknown, but we assume 1\% dust-to-gas ratio of
the Milky Way as an example, i.e. ${\cal D}=10^{-4}$. Note that the
results obtained in the following discussions are linearly scaled by the
value of ${\cal D}$. The mean hydrogen density in the Universe at
$z\sim10$ is $3\times10^{-4}$ cm$^{-3}$. Table 2 is a summary of the
assumed quantities and results obtained below.

\begin{table}[t]
 \renewcommand{\arraystretch}{1.2}
 \vspace{-.3cm}
 \caption{Photoelectric heating in the early Universe.}
 \vspace{-.1cm}
 \begin{center}
  \begin{tabular}{lc} \hline
   \multicolumn{2}{l}{Common setting} \\ \hline
   $z$ & 10 \\
   $n_{\rm H}$ & $3\times10^{-4}$ cm$^{-3}$ \\
   $\cal D$ & $10^{-4}$ \\
   size distribution & MRN \\
   $J_{\rm L}$ & $1\times10^{-21}$ erg s$^{-1}$ cm$^{-2}$ 
       Hz$^{-1}$ sr$^{-1}$ \\
   $p$ & 1 \\
   $E^{\rm max}_{\rm UV}$ & 13.6 eV \\ \hline
   \multicolumn{2}{l}{Nonionizing UV only} \\ \hline
   $T$ & 30 K \\
   $x_{\rm e}$ & $10^{-4}$ \\
   $\Gamma_{\rm pe}$ & $7\times10^{-36}$ erg s$^{-1}$ cm$^{-3}$ \\ 
   $t_{\rm pe}$ & $9\times10^9$ yr \\ \hline
   \multicolumn{2}{l}{With X-ray background} \\ \hline
   $E_{\rm X}^{\rm min}$ & 300 eV \\
   $T$ & $10^4$ K \\
   $x_{\rm e}$ & $0.3$ \\
   $\Gamma_{\rm pe}$ & $2\times10^{-33}$ erg s$^{-1}$ cm$^{-3}$ \\ 
   $\Gamma_{\rm pi,X}^{\rm HI}$ &  $2\times10^{-30}$ 
       erg s$^{-1}$ cm$^{-3}$\\ \hline
  \end{tabular}
 \end{center}
\end{table} 

In the nonionizing UV radiation only case, there is no efficient heating
mechanism for the whole of the Universe although primordial
objects can heat up their surrounding gas locally. Thus, the temperature
of the gas far away the sources is kept to be that of the cosmic
background radiation at the epoch: $\sim30$ K. The electron fraction
$x_{\rm e}$, i.e. the number density of electron relative to that of
hydrogen nucleus, is $\sim10^{-4}$ in this low temperature IGM (Galli \&
Palla 1998). The nonionizing UV photons still cause the photoelectric
effect of grains. In the assumed setting, we have found that grains are
positively charged and the dust photoelectric heating rate becomes 
$\Gamma_{\rm pe}\simeq7\times10^{-36}$ erg s$^{-1}$ cm$^{-3}$ for the
MRN size distribution with a graphite and silicate mixture (50\%
each in mass). We compare this heating rate with the gas thermal energy
density: $U_{\rm gas}=(3/2)n_{\rm H} k_{\rm B} T$. The time-scale
doubling the gas temperature with the photoelectric heating is given by 
$t_{\rm pe}\equiv U_{\rm gas}/\Gamma_{\rm pe}\simeq9\times10^9$ yr
Since the age of the Universe at $z=10$ is about $5\times10^8$ yr, 
we conclude that the dust photoelectric heating is not very efficient in
this case although it may be the strongest heating mechanism for the
IGM. 

In the case with the additional X-ray background radiation, the IGM is
partially ionized by the X-ray and the temperature becomes $\sim10^4$ K
(e.g., Venkatesan et al.~2001). If we assume the ionization equilibrium
and optically thin for the X-ray, the electron fraction becomes 
$x_{\rm e}\simeq0.3$ for the current setting of the X-ray background.
In this medium, the grains are positively charged and the dust
photoelectric heating rate becomes 
$\Gamma_{\rm pe}\simeq2\times10^{-33}$ erg s$^{-1}$ cm$^{-3}$. We have
assumed the MRN size distribution with a graphite and silicate
mixture (50\% each in mass), again. However, the HI photoionization
heating is much more efficient as 
$\Gamma^{\rm HI}_{\rm pi,X}\simeq2\times10^{-30}$ 
erg s$^{-1}$ cm$^{-3}$.
Therefore, we again conclude that the dust photoelectric heating is
negligible in the early Universe filled with an X-ray background
radiation.

\section{Conclusion}

We have updated our calculations made in Inoue \& Kamaya (2003,2004) of
the dust photoelectric heating in the IGM with the new model of the dust
photoelectric yield by Weingartner et al.~(2006). This new yield model
takes into account the effect of the photon and electron transfer in
a grain, the photoelectric emission from inner shells of grain
constituent atoms, the Auger electron emission, and the secondary
electron emission. A comparison with the previous yield model by
Weingartner \& Draine (2001) show that the new yield is smaller than the
old one for $\sim100$ eV photons. This reduction of the yield is due to
the photon/electron transfer effect, and reduces the electric potential
on grains and the heating rate significantly. For example, if we
integrate over the grain size with the standard MRN distribution, the
dust photoelectric heating rate with the new yield model is a factor of
2--4 smaller than that with the old yield model. The photoelectric
heating rate is more important in lower density medium. If the
dust-to-gas ratio in the IGM is 1\% of that in the Milky Way and the
size distribution is the standard MRN model, the dust heating rate
dominates the HI and HeII photoionization heating rates when the gas
number density is less than $\sim10^{-6}$ cm$^{-3}$ even with the new
yield model.

We have examined the effect of the size distribution function on the
heating rate because the heating rate is inversely proportional to the
typical grain size as found in equation (11). Bianchi \& Ferrara (2005)
suggested that the size of the intergalactic dust is larger than 
$\sim0.1$ $\mu$m because smaller grains are destroyed by sputtering in
the hot gas halo during the transport of grains from the parent galaxy
to the IGM. In this case, the heating rate is reduced by a
factor of $\sim5$ relative to that with the standard MRN size
distribution. The size distributions expected by the dust formation
model in supernova ejecta are also examined. The heating rate with
the size distribution of the grains just produced in the ejecta is very
similar to that with the MRN distribution. In contrast, the heating rate
with the size distribution of the grains processed by the reverse shock
in the supernova remnant is a factor of $\sim3$ smaller than that with
the MRN model. The shock processed grains have larger size than the
pristine ones because smaller grains are destroyed. On the other hand,
if we put only small grains in the IGM, the heating rate increases
significantly. Therefore, we conclude that the size distribution of
grains in the IGM is an essential parameter for determining the dust
heating efficiency. Even in the worst case considered here, the dust
heating is expected to be the dominant heating mechanism in the IGM at
$z=0$ if the dust-to-gas ratio in the IGM is 1\% of that in the Milky
Way.

Since the dust photoelectric heating rate with the new yield model is
reduced by a factor of 2--4 relative to that with the old yield model,
the upper limit on the amount of the intergalactic dust obtained by
Inoue \& Kamaya (2003,2004) may be affected. Indeed, the limit based on
the thermal history of the IGM should be raised by a few factor. 
However, their final upper limit is mainly obtained from the reddening
measurements of $z=0.5$ supernovae Ia. Therefore, their conclusion would
not be affected very much.

Bolton et al.~(2008) suggested an inverted temperature--density relation
in the lower density IGM at $z=2$--3 based on recent observations of the
Lyman $\alpha$ forest in QSOs' spectra. To explain this interesting
phenomenon, we need a heating mechanism more efficient in a lower
density medium. The dust photoelectric heating has such a property. 
Indeed, the dust heating rate even with the new yield model is a factor
of 2 larger than the HeII photoionization heating rate in a medium with a
density of 1/10 of the mean in the Universe at $z=2$ if the dust-to-gas
ratio is 1\% of that in the Milky Way. Thus, the possibility of the dust
heating is worth examining more in detail. For this aim, the simple
formula describing the dust photoelectric heating in the IGM presented
in equation (16) will be very useful.

Finally, we have discussed the effect of the dust photoelectric heating
in the early Universe. Before the cosmic reionization, the ionizing
background radiation is not established, but there may be 
nonionizing UV background and X-ray background radiations. In the low
temperature IGM only with a nonionizing UV background radiation, the
dust photoelectric heating is not very efficient although it may be the
strongest heating mechanism in the medium. In the partially ionized IGM
with an X-ray background radiation, the HI photoionization heating rate
is three orders of magnitude larger than the dust heating rate if the
dust-to-gas ratio is 1\% of that in the Milky Way. Therefore, we
conclude that the dust photoelectric heating in the early Universe is
not very important at least in the mean density environment.

\vspace{1cm}
We would appreciate comments from the reviewers, 
B.~T.~Draine and M.~M.~Abbas, which improve the quality of this paper
very much. 
We are grateful to the conveners of the session ``Cosmic Dust'' 
in the 5th annual meeting of the Asia-Oceania Geosciences Society 
for organizing the interesting workshop. 
AKI is also grateful to all members of the Department of Physics, 
Nagoya University, especially the $\Omega$ Laboratory led by Tsutomu T.\
Takeuchi, for their hospitality during this work. AKI is supported by
KAKENHI (the Grant-in-Aid for Young Scientists B: 19740108) by The
Ministry of Education, Culture, Sports, Science and Technology (MEXT) of
Japan.



\section*{References}
\small

Abbas, M. M., et al., Photoelectric Emission Measurements on the Analogs
 of Individual Cosmic Dust Grains, \textit{The Astrophys. J.}, 
 \textbf{645}, 324--336, 2006.

Abel, T., Haehnelt, M. G., 
Radiative Transfer Effects during Photoheating of the Intergalactic Medium, 
\textit{The Astrophys. J. Letters}, \textbf{520}, L13--L16, 2001.

Aguirre, A., Haiman, Z., Cosmological Constant or Intergalactic Dust?
 Constraints from the Cosmic Far-Infrared Background, 
\textit{The Astrophys. J.}, \textbf{532}, 28--36, 2000.

Aguirre, A., Hernquist, L., Katz, N., Gardner, J., Weinberg, D.,
Enrichment of the Intergalactic Medium by Radiation Pressure-driven Dust Efflux,
\textit{The Astrophys. J. Letters}, \textbf{556}, L11--L14, 2001a. 

Aguirre, A., Hernquist, L., Schaye, J., Weinberg, D. H., Katz, N.,
 Gardner, J., Metal Enrichment of the Intergalactic Medium at $z=3$ by
 Galactic Winds, \textit{The Astrophys. J.}, \textbf{560}, 599--605, 2001b.

Aguirre, A., Hernquist, L., Schaye, J., Katz, N., Weinberg, D. H.,
 Gardner, J., Metal Enrichment of the Intergalactic Medium in
 Cosmological Simulations, \textit{The Astrophys. J.}, \textbf{561},
 521--549, 2001c.

Alton, P. B., Davies, J. I., Bianchi, S., 
Dust outflows from starburst galaxies, 
\textit{M.N.R.A.S.}, \textbf{343}, 51--63, 1999.

Bendo, G. J., et al.,
The Spectral Energy Distribution of Dust Emission in the Edge-on Spiral
 Galaxy NGC 4631 as Seen with Spitzer and the James Clerk Maxwell
 Telescope, \textit{The Astrophys. J.}, \textbf{652}, 283--305, 2006.

Bertoldi, F., Carilli, C. L., Cox, P., Fan, X., Strauss, M. A., Beelen,
 A., Omont, A., Zylka, R., Dust emission from the most distant quasars, 
\textit{Astronomy and Astrophysics}, \textbf{406}, L55--L58, 2003.

Bianchi, S., Ferrara, A.,
Intergalactic medium metal enrichment through dust sputtering,
\textit{M.N.R.A.S.}, \textbf{358}, 379--396, 2005.

Bolton, J. S., Viel, M., Kim, T.-S., Haehnelt, M. G., Carswell, R. F.,
Possible evidence for an inverted temperature-density relation in the
 intergalactic medium from the flux distribution of the Ly$\alpha$ forest, 
\textit{M.N.R.A.S.}, \textbf{386}, 1131--1144, 2008.

Corasaniti, P. S., The impact of cosmic dust on supernova cosmology
\textit{M.N.R.A.S.}, \textbf{372}, 191--198, 2006.

Draine, B. T., Photoelectric heating of interstellar gas, 
\textit{Astrophysical Journal Supplement Series}, \textbf{36}, 
595--619, 1978

Draine, B. T., 
Evolution of interstellar dust, in \textit{The evolution of the
 interstellar medium}, Edited by L. Blitz, 193--205, 
 Astronomical Society of the Pacific, San Francisco, 1990.

Draine, B. T., 
Scattering by Interstellar Dust Grains. I. Optical and Ultraviolet,
\textit{The Astrophys. J.}, \textbf{598}, 1017--1025, 2003.

Draine, B. T., Salpeter, E. E., 
Destruction mechanisms for interstellar dust, 
\textit{The Astrophys. J.}, \textbf{231}, 438--455, 1979.

Draine, B. T., Sutin, B.,
Collisional charging of interstellar grains, 
\textit{The Astrophys. J.}, \textbf{320}, 803--817, 1987.

Dwek, E., The Evolution of the Elemental Abundances in the Gas and Dust
 Phases of the Galaxy, \textit{The Astrophys. J.}, \textbf{501}, 
 645--665, 1998.

Elfgren, E., D{\'e}sert, F.-X., Guiderdoni, B.,
Dust distribution during reionization, 
\textit{Astronomy and Astrophysics}, \textbf{476}, 1145--1150, 2007.

Fan, X., Carilli, C. L., Keating, B., 
Observational Constraints on Cosmic Reionization, 
\textit{Annual Review of Astronomy \& Astrophysics}, \textbf{44}, 
415--462, 2006.

Ferrarotti, A. S., Gail, H.-P., 
Composition and quantities of dust produced by AGB-stars and returned to
 the interstellar medium 
\textit{Astronomy and Astrophysics}, \textbf{447}, 553

Ferrara, A., Ferrini, F., Barsella, B., Franco, J.,
Evolution of dust grains through a hot gaseous halo,
\textit{The Astrophys. J.}, \textbf{381}, 137--146, 1991.

Ferrara, A., Pettini, M., Shchekinov, Yu. A.,
Mixing metals in the early Universe,
\textit{M.N.R.A.S.}, \textbf{319}, 539--548, 2000.

Fruchter, A., Krolik, J. H., Rhoads, J. E.,
X-Ray Destruction of Dust along the Line of Sight to $\gamma$-Ray Bursts,
\textit{The Astrophys. J.}, \textbf{563}, 597--610, 2001.

Galli, D., Palla, F., The chemistry of the early Universe,
\textit{Astronomy and Astrophysics}, \textbf{335}, 403--420, 1998.

Goobar, A., Bergstr{\"o}m, L., M{\"o}rtsell, E., Measuring the
 properties of extragalactic dust and implications for the Hubble
 diagram, \textit{Astronomy and Astrophysics}, \textbf{384}, 1--10, 2002.

Gunn, J. E., Peterson, B. A., 
On the Density of Neutral Hydrogen in Intergalactic Space,
\textit{The Astrophys. J.}, \textbf{142}, 1633--1641, 1965.

Haardt, F., Madau, P.,
Radiative Transfer in a Clumpy Universe. II. The Ultraviolet
 Extragalactic Background, 
\textit{The Astrophys. J.}, \textbf{461}, 20--37, 1996.

Hirashita, H., Nozawa, T., Kozasa, T., Ishii, T. T., Takeuchi, T. T., 
Extinction curves expected in young galaxies,
\textit{M.N.R.A.S.}, \textbf{357}, 1077--1087, 2005.

Hirashita, H., Nozawa, T., Takeuchi, T. T., Kozasa, T., 
Extinction curves flattened by reverse shocks in supernovae,
\textit{M.N.R.A.S.}, \textbf{384}, 1725--1732, 2005.

Hui, L., Gnedin, N. Y.,
Equation of state of the photoionized intergalactic medium,
\textit{M.N.R.A.S.}, \textbf{292}, 27--42, 1997.

Inoue, A. K., and Kamaya, H., 
Constraint on intergalactic dust from thermal history of intergalactic
 medium, \textit{M.N.R.A.S.}, \textbf{341}, L7--L11, 2003. 

Inoue, A. K., and Kamaya, H., Amount of intergalactic dust: constraints
 from distant supernovae and the thermal history of the intergalactic
 medium, 
\textit{M.N.R.A.S.}, \textbf{350}, 729--744, 2004.

Jones, A. P., Tielens, A. G. G. M., Hollenbach, D. J., 
Grain Shattering in Shocks: The Interstellar Grain Size Distribution, 
\textit{The Astrophys. J.}, \textbf{469}, 740--764, 1996.

Loeb, A., Barkana, R.,
The Reionization of the Universe by the First Stars and Quasars,
\textit{Annual Review of Astronomy and Astrophys.}, \textbf{39}, 19--66,
 2001.

Mathis, J. S., Rumpl, W., and Nordsieck, K. H., 
 The size distribution of interstellar grains, 
 \textit{Astrophys. J.}, \textbf{217}, 425--433, 1977.

Maiolino, R., Schneider, R., Oliva, E., Bianchi, S., Ferrara, A.,
 Mannucci, F., Pedani, M., Roca Sogorb, M, 
A supernova origin for dust in a high-redshift quasar, 
\textit{Nature}, \textbf{431}, 533--535, 2004.

Montier, L. A., Giard, M.,
The importance of dust in cooling and heating the InterGalactic Medium, 
\textit{Astronomy and Astrophysics}, \textbf{417}, 401--409, 2004.

Nath, B. B., Sethi, S. K., and Shchekinov, Y., 
Photoelectric heating for dust grains at high redshifts, 
 \textit{M.N.R.A.S.}, \textbf{303}, 1--14, 1999.

Nozawa,, T. Kozasa, T., Umeda, H., Maeda, K., Nomoto, K.,
Dust in the Early Universe: Dust Formation in the Ejecta of Population
 III Supernovae, \textit{The Astrophys. J.}, \textbf{598}, 785--803, 2003.

Nozawa,, T. Kozasa, Habe, A., Dwek, E., Umeda, H., Tominaga, N., Maeda,
 K., Nomoto, K., Evolution of Dust in Primordial Supernova Remnants: Can
 Dust Grains Formed in the Ejecta Survive and Be Injected into the Early
 Interstellar Medium$?$, \textit{The Astrophys. J.}, \textbf{598},
 955--966, 2007.

Osterbrock, D. E., Ferland, G. J., in \textit{Astrophysics of Gaseous
 Nebulae and Active Galactic Nuclei Second Edition}, University Science
 Books, SauSalito California, 2006.

Rho, J., Kozasa, T., Reach, W.\ T., Smith, J.\ D., Rudnick, L., DeLaney,
 T., Ennis, J.\ A., Gomez, H., Tappe, A.,
Freshly Formed Dust in the Cassiopeia A Supernova Remnant as Revealed by
 the Spitzer Space Telescope,
\textit{The Astrophys. J.}, \textbf{673}, 271--282, 2008.

Scott, J., Bechtold, J., Dobrzycki, A., Kulkarni, V. P., 
A Uniform Analysis of the Ly$\alpha$ Forest at $z=0-5$. II. Measuring
 the Mean Intensity of the Extragalactic Ionizing Background Using the
 Proximity Effect, \textit{The Astrophys. J. Supp.}, \textbf{130},
 67--89, 2000. 

Spizer, L. Jr.,
The Dynamics of the Interstellar Medium. II. Radiation Pressure,
\textit{The Astrophys. J.}, \textbf{94}, 232--244, 1941.

Venkatesan, A., Giroux, M. L., Shull, J. M.,
Heating and Ionization of the Intergalactic Medium by an Early X-Ray
 Background, \textit{The Astrophys. J.}, \textbf{563}, 1--8, 2001.

Weingartner, J., and Draine, B. T., Photoelectric Emission from
 Interstellar Dust: Grain Charging and Gas Heating, 
 \textit{The Astrophys. J. Suppl.}, \textbf{134}, 263--281, 2001.

Weingartner, J., Draine, B. T., and Barr, D. K., Photoelectric Emission
 from Dust Grains Exposed to Extreme Ultraviolet and X-Ray Radiation, 
 \textit{The Astrophys. J.}, \textbf{645}, 1188--1197, 2006.

Williams, B. J., et al.,
Dust Destruction in Fast Shocks of Core-Collapse Supernova Remnants in
 the Large Magellanic Cloud, \textit{The Astrophys. J. Letters},
 \textbf{652}, L33--L56, 2006.

Xilouris, E., Alton, P., Alikakos, J., Xilouris, K., Boumis, P., Goudis, C.,
Abundant Dust Found in Intergalactic Space,
\textit{The Astrophys. J. Letters}, \textbf{651}, L107--L110, 2008.

Yamada, K., Kitayama, T., Infrared Emission from Intracluster Dust
 Grains and Constraints on Dust Properties, 
 \textit{Publications of the Astronomical Society of Japan}, 
 \textbf{57}, 611--619, 2005.

Yun, M. S., Ho, P. T. P., Lo, K. Y., A High-Resolution Image of Atomic
 Hydrogen in the M81 Group of Galaxies, \textit{Nature}, \textbf{372}, 
 530--532, 1994.

Zhang, P., Corasaniti, P. S., Cosmic Dust Induced Flux Fluctuations: Bad
 and Good Aspects, \textit{The Astrophys. J.}, \textbf{657}, 71--75, 
 2007.


\end{document}